# M. P. Korkina[1], O. O. Iegurnov[2]

*Oles' Honchar Dnipropetrovs'k National University*

# MATCHING OF STEPHANI AND DE SITTER SOLUTIONS ON THE HYPERSURFACE OF CONSTANT TIME

Spherically symmetric solution for perfect fluid with homogeneous density and inhomogeneous pressure has been considered. This solution is well known as Stephani solution and it has been investigated in numerous papers. It has been reobtained by mass-function method. Also the meaning of arbitrary functions, which are present in the solution, has been discussed. The matching of this solution and de Sitter solution has been done on a hypersurface of constant time. The matching has been done with the Lichnerowicz – Darmois conditions. The coordinates of the de Sitter solution have been taken in a general form as arbitrary functions that depend on the Stephani's time and radial coordinate. The matching has been done as for partial cases (flat, open, closed) and for general case, that not concretize type of curvature. An equality of the densities and a bound of the pressures have been shown on the matching hypersurface. Also, restrictions on arbitrary functions (coordinates of the de Sitter solution) have been found.

Keywords: perfect fluid, Stephani solution, de Sitter solution, Lichnerowicz - Darmois conditions.

Рассмотрено сферически симметричное решение для идеальной жидкости с однородной плотностью энергии и неоднородным давлением. Это решение – хорошо известное решение Стефани, которому исследовалось и обсуждалось в множестве статей. Это решение было получено повторно с помощью метода массовой функции. Также обсуждается смысл произвольных функций, которые присутствуют в решении. Выполнена сшивка этого решения с решением де Ситтера по гиперповерхности постоянного времени. Сшивка была выполнена с помощью условий Лихнеровича-Дарма. Координаты решения де Ситтера были взяты в общем виде, как функции от координаты времени и пространственной радиальной координаты решения Стефани. Сшивка была выполнена для частных случаев (плоский, закрытый, открытый мир) и для общего случая, в котором не конкретизируется кривизна. Показано, что на гиперповерхности сшивки наблюдается равенство плотностей энергий и скачок давления. Также найдены ограничения на произвольные функции – координаты решения де Ситтера.

Ключевые слова: идеальная жидкость, решение Стефани, решение де Ситтера, условия Лихнеровича – Дарма.

Розглянуто сферично симетричний розв'язок для ідеальної рідини з однорідною густиною енергії та неоднорідним тиском. Цей розв'язок  - добре відомий розв'язок  Стефані, який досліджувався та обговорювався у багатьох статтях. Цей розв'язок цікавий тим, що він може бути використаний для побудови космологічних моделей. Цей розв'язок було отримано повторно за допомогою метода масової функції. Також обговорюється значення довільних функцій, що присутні у розв'язку. Зроблена сшивка цього розв'язку з розв'язком де Сіттера за гіперповерхнею постійного часу. Сшивка була зроблена за допомогою умов Ліхнеровіча-Дармуа. Координати для розв'язку де Сіттера було обрано у загальному вигляді, як довільні функції від координати часу та просторової радіальної координати розв'язку Стефані. Сшивка була зроблена для часткових випадків (плоский, закритий, відкритий) та для загального, у якому не конкретизується кривизна. Показано, що на гіперповерхні сшивки спостерігається рівність густин енергій та стрибок тиску. Також знайдені обмеження на довільні функції – координати розв'язку де Сіттера.

Ключові слова: ідеальна рідина, розв'язок Стефані, розв'язок де Сіттера, умови Ліхнеровича – Дармуа.

[1] 958korkin@rambler.ru

[2] iegurnov@gmail.com

## Introduction

The most general class of non-static, perfect fluid solutions of Einstein's equations that are conformally flat is known as the "Stephani Universe" [1-5]. The spherically symmetric Stephany solution has been examined in numerous papers. A comprehensive review is presented in [5]. There are many papers devoted to applying this solution as star models, as generalization of the FLRW, as cosmological model [5-7]. In our opinion, this solution is attractive for cosmological model for many reasons. Firstly, it is shear-free and inhomogeneous. The absence of a shear makes it simple for the cosmological purpose. The assumption of homogeneity is just a first approximation introduced to simplify Einstein's equations. So far this assumption has worked well, but future and modern observations can not be precise without taking into account inhomogeneity. And due to the fact, that modern and future observation dates become more and more precisely and that the smallest deviations from the standard model can be detected with high level of accuracy soon, makes inhomogeneous models actual. Secondly, Stephani solution has general form in contradistinction to the FLRW solution, where three solutions (flat, open, closed), non-transforming into each other, exist. Thirdly, the spatial curvature of this solution depends on time only via an arbitrary function, this fact is discussed in [3, 8].

Physical interpretation of the Stephani Universe is obscure. It is due to the many arbitrary functions and peculiar inhomogeneity – inhomogeneity is contained in the pressure (depend on time and spatial coordinates), but density is homogeneous (depend on time only). It needs information in order to determine some arbitrary functions. May be the main reason to use this solution in cosmological modeling is the fact that it is generalization of the FLRW solution, and, in our opinion, studying more general solution is promising. This solution generalizes not only the FLRW but the de Sitter solution too [3]. In this connection, the idea to examine the Stephani solution on the de Sitter background looks reasonable.

In the first part of the article the solution for perfect fluid with inhomogeneous pressure (the Stephani solution) has been reobtained with mass-function method [9-12]. Some properties of it are discussed, in particular, we discuss a sense of the arbitrary functions and transformation to the FLRW and to the de Sitter solutions.

In the second part of the article the matching of the de Sitter and the Stephani solutions on the hypersurface of constant time is done. Lichnerowicz – Darmois conditions were used. Some consequences of matching are discussed.

## Mass-function method

Mass-function method essentially simplifies the appearance of the Einstein's equation in contrast to the standard one, it makes them easier for work. The mass-function was introduced in [9] and discussed in [10-12]. As shown in [9], the mass-function is invariant and in our consideration may be determined as full energy limited by some hypersurface of constant time and coordinates. For spherically symmetric metric:

$$dS^2 = e^{\nu(R,t)}dt^2 - e^{\lambda(R,t)}dR^2 - r^2(R,t)d\sigma^2, \tag{1}$$

where $d\sigma^2 = d\theta^2 - \sin^2(\theta)d\phi^2$, mass-function $m(R,t)$ is:

$$m(R,t) = r(R,t)(1 + e^\Phi - e^\Omega), \tag{2}$$

$$e^\Phi = e^{-\nu}\dot{r}^2(R,t), \quad e^\Omega = e^{-\lambda}r'^2(R,t), \tag{3}$$

where $\dot{r}(R,t) = \dfrac{\partial r(R,t)}{\partial t}, \ r'(R,t) = \dfrac{\partial r(R,t)}{\partial R}.$

Einstein's equations with mass-function have following form:

$$\begin{cases} m' = r^2 r' T_0^0, \\ \dot{m} = r^2 \dot{r} T_1^1, \\ 2\dot{m}' = \dot{m}\Phi' + m'\dot{\Omega} + 4r\dot{r}r'T_2^2, \\ 2\dot{r}' = \dot{r}\Phi' + r'\dot{\Omega}. \end{cases} \tag{4}$$

## Obtaining Stephani solution

This solutions was first found by Stephani [13] as a special example of a space-time embeddable in a flat five-dimensional space, and later reobtained by Krasinski [2]. We reobtained this solution with mass-function method.

The Stephani solution is an isotropic solution for perfect fluid with homogeneous density $\rho = \rho(t)$ and inhomogeneous pressure $p = p(R,t)$ (in spherically symmetric consideration). The stress-energy tensor for this perfect fluid is: $T_0^0 = \rho(t)$, $T_1^1 = T_2^2 = T_3^3 = -p(R,t)$. The Einstein's field equations become:

$$\begin{cases} m' = r^2 r' \rho, \\ \dot{m} = -r^2 \dot{r} p, \\ 2\dot{m}' = \dot{m}\Phi' + m'\dot{\Omega} - 4r\dot{r}r'p, \\ 2\dot{r}' = \dot{r}\Phi' + r'\dot{\Omega}. \end{cases} \tag{5}$$

Expressing the mass-function from the first equation of the set and the substituting it into the third one gives:

$$\frac{\dot{\rho}}{\rho}\left(\frac{1}{3}r\Phi' + \frac{4}{3}r' - 2r'\right) = 0, \tag{6}$$

from this equation expression for $\Phi$ is obtained:

$$\Phi = \ln r^2 \psi^2, \tag{7}$$

where $\psi = \psi(t)$ - arbitrary function of integration.
The fourth equation of the set (5) gives us the expression for $\Omega$:

$$\Omega = \ln \frac{r'^2}{r^2 k'^2}, \tag{8}$$

where $k = k(R)$ - arbitrary function (prime is used for convenience).
From (1) and (3) the metric is obtained:

$$dS^2 = \frac{\dot{r}^2}{r^2 \psi^2} dt^2 - r^2 (k'^2 dR^2 + d\sigma^2). \tag{9}$$

With (7) and (8) the mass-function is obtained:

$$m = r(1 + r^2 \psi^2 - \frac{r'^2}{r^2 k'^2}). \tag{10}$$

The expression (10) with the first equation of (1.5) gives us:

$$\rho \frac{r^3}{3} = r + r^3 \psi^2 - \frac{r'^2}{rk'^2}, \tag{11}$$

it is integrated in elementary functions providing the expressing for $r = r(R,t)$:

$$r(R,t) = 2(e^{k(R)+\eta(t)} - \zeta(t)e^{-k(R)-\eta(t)})^{-1}, \tag{12}$$

$$\zeta(t) = \psi^2(t) - \frac{1}{3}\rho(t), \tag{13}$$

and $\eta(t)$ is arbitrary function of integration.
Depending on the sign of $\zeta(t)$ the expression (12) is:

$$\begin{aligned} r^{-1} &= \sqrt{\zeta} \cdot \sinh(k + \alpha), \ \zeta > 0, \\ r^{-1} &= \sqrt{|\zeta|} \cdot \cosh(k + \alpha), \ \zeta < 0, \\ r^{-1} &= e^{k+\alpha}, \ \zeta = 0, \end{aligned} \tag{14}$$

where $e^\eta = \sqrt{|\zeta|} \cdot e^\alpha$.

In contrast to the FLRW solution, where there are three non-transforming into each other solutions (flat, closed, open), there is a general solution here with flat, open, closed solutions as special cases. The existence of this solution shows that the distinction between the closed and open Universe is not required by Einstein's theory of gravitation as such, but is due to the very strong symmetry assumptions, which are put into the models right from the beginning. From (5) it is also possible to obtain equation that links density and pressure:

$$p(R,t) = -\rho(t) - \frac{\dot{\rho}}{3} \frac{r(R,t)}{\dot{r}(R,t)}. \tag{15}$$

### The FLRW and the de Sitter solutions as special cases of the Stephani solution

Stephani solution, as mentioned above, is a generalization of the FLRW solution and the de Sitter one. When $\zeta(t) = 0$, the Stephani solution is transformed to the flat FLRW solution. If $\zeta(t) \neq 0$ transformation occurs in such a way:

$$\dot{\eta}(t) = 0, \psi(t) = \frac{\dot{a}(t)}{a(t)},$$

$$\zeta(t) = -\frac{1}{a^2(t)} : k(R) = \ln\cot\frac{R}{2}, \tag{16}$$

$$\zeta(t) = \frac{1}{a^2(t)} : k(R) = \ln\coth\frac{R}{2}.$$

A density of the de Sitter solution is $\rho = \frac{1}{a_\lambda^2} = const.$ One may be obtained from (13, 16):

$$\zeta(t) = 0 : a(t) = e^{\frac{t}{a_\lambda}},$$

$$\zeta(t) = -\frac{1}{a^2(t)} : a(t) = a_\lambda \cosh\frac{t}{a_\lambda}, \tag{17}$$

$$\zeta(t) = \frac{1}{a^2(t)} : a(t) = a_\lambda \sinh\frac{t}{a_\lambda}.$$

### Arbitrary functions and their sense

The Stephani solution contains four arbitrary functions: $k(R), \zeta(t), \psi(t), \eta(t)$. Also in our consideration we don't set the equation of state, i.e. $\rho(t)$ undefined. It may appear that determination of arbitrary functions is elusive. It is true, but analysis of solution (9) show, that it is possible to understand their meaning.

The function $k(R)$ may be chosen arbitrary, because it leads to a transition to another coordinate system, only. The part $\left(\frac{dk}{dR}\right)^2 dR^2 = dk^2$ in expression (9) is just transformation from $R$ to $k$. The coordinate transformation may be chosen in such way, that spatial part of the solution (9) is conformal to one of the three homogeneous and isotropic spaces:

$$dk^2 + d\sigma^2 = \frac{1}{\sinh^2(R_1)}\left(dR_1^2 + \sinh^2(R_1)d\sigma^2\right) =$$

$$= \frac{1}{\sin^2(R_2)}\left(dR_2^2 + \sin^2(R_2)d\sigma^2\right) = \frac{1}{R_3^2}\left(dR_3^2 + R_3^2 d\sigma^2\right). \tag{18}$$

Chosen $\eta(t)$ is also referring to the coordinate transformation. Thus (9) takes the form:

$$dS^2 = \frac{r_\eta^2}{r^2\psi^2}d\eta^2 - r^2(k'^2 dR^2 + d\sigma^2) \tag{19}$$

where $r_\eta = \frac{\partial r}{\partial \eta}$.

The analysis of invariants of the spatial curvature tensor of the metric (9) showed that invariants depend on the arbitrary function $\zeta(t)$, only. The scalar curvature tensor and the Kretschmann scalar, for example, are:

$$R = 6\zeta(t),$$
$$R_{\mu\nu\lambda\sigma}R^{\mu\nu\lambda\sigma} = 12\zeta^2(t). \tag{20}$$

Thereby spatial curvature depends only by $\zeta(t)$. The type of space (flat, open, closed) is determined by a sign of $\zeta(t)$. So it is possible to assume that $\zeta(t)$ completely determines spatial curvature. This fact is also discussed in [3, 8].

Thus, $\psi(t)$ makes sense of the critical energy density. When spatial curvature is zero, then $\zeta(t) = 0$ and from (13):

$$\psi^2(t) = \frac{1}{3}\rho_c(t) \tag{21}$$

### Lichnerowicz - Darmois conditions

The Lichnerowicz-Darmois matching conditions [14] are two metrics

$$dS_1^2 = g_{\mu'\nu'}dx^{\mu'}dx^{\nu'},$$
$$dS_2^2 = g_{\mu\nu}dx^{\mu}dx^{\nu}, \tag{22}$$

are said to match across some hypersurface if first and second fundamental forms of this hypersurface are identical for the two metrics.

First fundamental form:

$$dl_1^2 = a_{ik}du^i du^k \quad (i,k = 1,2,3), \tag{23}$$

where $a_{ik} = g_{\mu\nu}\xi_i^{\mu}\xi_k^{\nu}$, $\xi_i^{\mu} = \dfrac{\partial x^{\mu}}{\partial x^i}$.

Second fundamental form:

$$dl_2^2 = b_{ik}du^i du^k, \tag{24}$$

where $b_{ik} = \nu_{\mu;\nu}\xi_i^{\mu}\xi_k^{\nu}$, $\xi_i^{\mu}$ are tangent vectors to the hypersurface, $\nu^{\mu}$ are normal vectors.

### Matching Stephani and de Sitter solutions in general case

Matching has been done on the hypersurface of Stephani's constant time. Time and spatial coordinate of the de Sitter solution have been chosen as arbitrary functions of the Stephani's time and radial coordinate. Both metrics has been taken in general form without more precise definition of their curvature.

The Stephani metric, as mentioned above, is:

$$dS_{st}^2 = \frac{\dot{r}^2}{r^2\psi^2}d\tau^2 - r^2(dk^2 + d\sigma^2), \tag{25}$$

where $r = r(k,\tau)$, $\psi = \psi(\tau)$, $\dot{r} = \dfrac{\partial r}{\partial \tau}$.

The matching is performed on the hypersurface $\tau = const.$ $\xi_i^{\mu}$ on this hypersurface is:

$$\xi_i^{\mu} = \begin{pmatrix} 0 & 0 & 0 & 0 \\ 0 & 1 & 0 & 0 \\ 0 & 0 & 1 & 0 \\ 0 & 0 & 0 & 1 \end{pmatrix}. \tag{26}$$

Normal vectors are found from equation:

$$\begin{cases} \nu_{\mu}\nu^{\mu} = 1, \\ \nu_{\mu}\xi_i^{\mu} = 0. \end{cases} \tag{27}$$

Normal vector has only one nonzero component:

$$\nu_0 = \frac{\dot{r}}{r\psi}. \tag{28}$$

First and second fundamental forms for the Stephani solution are:

$$dl_{st1}^2 = r^2 dk^2 + r^2 d\sigma^2, \tag{29}$$

$$dl_{st2}^2 = r^2 \psi dk^2 + r^2 \psi d\sigma^2. \tag{30}$$

The de Sitter metric:

$$dS_{ds}^2 = \left(1 - \frac{r_s^2}{a_\lambda^2}\right) dt^2 - \frac{1}{1 - \frac{r_s^2}{a_\lambda^2}} dr_s^2 - r_s^2 d\sigma^2, \tag{31}$$

$r_s = r_s(k,\tau)$, $t = t(k,\tau)$, $a_\lambda = \sqrt{\dfrac{3}{\Lambda}}$, $\Lambda$ is cosmological constant. We do the same manipulations with the de Sitter metric :

$$\xi_i^\mu = \begin{pmatrix} 0 & t' & 0 & 0 \\ 0 & r_s' & 0 & 0 \\ 0 & 0 & 1 & 0 \\ 0 & 0 & 0 & 1 \end{pmatrix}. \tag{32}$$

According to (2.6):

$$\nu_0 = \frac{r_s'}{\sqrt{A^{-1} r_s'^2 - A t'^2}}, \tag{33}$$

$$\nu_1 = \frac{t'}{\sqrt{A^{-1} r_s'^2 - A t'^2}}. \tag{34}$$

where the prime is a derivative for $k$, $A = 1 - \dfrac{r_s^2}{a_\lambda^2}$.

The first and second fundamental forms for the de Sitter solution are:

$$dl_{ds1}^2 = (A^{-1} r_s'^2 - A t'^2) dk^2 + r_s^2 d\sigma^2, \tag{35}$$

$$dl_{ds2}^2 = \frac{r_s t'(t'^2 A - 3 r_s'^2 A^{-1})}{a_\lambda^2 \sqrt{A^{-1} r_s'^2 - A t'^2}} dk^2 + \frac{r_s A t'}{\sqrt{A^{-1} r_s'^2 - A t'^2}} d\sigma^2. \tag{36}$$

From the equality of the first and the second fundamental forms the following equations can been obtain:

$$r^2 = r_s^2, \tag{37}$$

$$r^2 = A^{-1} r_s'^2 - A t'^2, \tag{38}$$

$$r^2 \psi = \frac{r_s t'(t'^2 A - 3 r_s'^2 A^{-1})}{\sqrt{A^{-1} r_s'^2 - A t'^2}}, \tag{39}$$

$$r^2 \psi = \frac{r_s A t'}{\sqrt{A^{-1} r_s'^2 - A t'^2}}. \tag{40}$$

From these matching conditions equality of the energy densities on the hypersurface $\tau = const.$ follow. From (39) and (40) we have

$$r_s A t' = \frac{r_s}{a_\lambda^2} t'(t'^2 A - 3 r_s'^2 A^{-1}), \tag{41}$$

and with (38)

$$A = \frac{1}{a_\lambda^2}(-r_s^2 - 2r_s'^2 A^{-1}),$$

$$r_s'^2 = \frac{r_s^2 - a_\lambda^2}{2},$$

with (37) and (32) we obtained

$$\rho_{st} = \frac{1}{a_\lambda^2}. \tag{42}$$

But this is de Sitter's energy density. So, on the hypersourface $\tau = const.$ the equality of the energy densities holds:

$$\rho_{st} = \rho_{ds}. \tag{43}$$

**Matching Stephani and de Sitter solutions in the flat case**

De Sitter metric for the flat case has the form:

$$dS_{ds}^2 = dt^2 - a_\lambda^2 e^{\frac{2t}{a_\lambda}}(dr_s^2 + r_s^2 d\sigma^2). \tag{44}$$

After transformation its spatial part to the convenient form it takes the form:

$$dS_{ds}^2 = dt^2 - a_\lambda^2 e^{\frac{2t}{a_\lambda} + 2X}(dX^2 + d\sigma^2). \tag{45}$$

The Stephani metric is:

$$dS_{st}^2 = \frac{1}{\psi^2}d\tau^2 - 4e^{-2(R-\tau)}(dR^2 + d\sigma^2). \tag{46}$$

As mentioned, $t = t(\tau, R)$, $X = X(\tau, R)$. Dots and primes mean derivatives with respect to the time and the radial coordinate, respectively. For the Stephani solution the normal vector has one non zero component on the matching hypersurface $\tau = const.$:

$$\nu_0 = \frac{1}{\psi}. \tag{47}$$

For the de Sitter solution normal vector has two non-zero components:

$$\nu_0 = \frac{X' a_\lambda e^{\frac{t}{a_\lambda} + X}}{\sqrt{X'^2 a_\lambda^2 e^{2(\frac{t}{a_\lambda} + X)} - t'^2}}, \tag{48}$$

$$\nu_1 = -\frac{t' a_\lambda e^{\frac{t}{a_\lambda} + X}}{\sqrt{X'^2 a_\lambda^2 e^{2(\frac{t}{a_\lambda} + X)} - t'^2}}, \tag{49}$$

From the equality of the first fundamental forms, two matching conditions can be obtained:

$$4e^{-2(R+\tau)} = X'^2 a_\lambda^2 e^{2(\frac{t}{a_\lambda} + X)} - t'^2, \tag{50}$$

$$4e^{-2(R+\tau)} = a_\lambda^2 e^{2(\frac{t}{a_\lambda} + X)}(X'^2 - 1). \tag{51}$$

Right-hand sides of these equations are equal, from this equality we obtain:

$$t'^2 = a_\lambda^2 e^{2(\frac{t}{a_\lambda} + X)}(X'^2 - 1). \tag{52}$$

From the equality of the second fundamental forms, such two matching conditions follow:

$$4\psi e^{-2(R+\tau)} = -\frac{a_\lambda e^{\frac{t}{a_\lambda} + X}}{\sqrt{X'^2 a_\lambda^2 e^{2(\frac{t}{a_\lambda} + X)} - t'^2}}\left(a_\lambda X' e^{2(\frac{t}{a_\lambda} + X)} + t'\right), \tag{53}$$

$$4\psi e^{-2(R+\tau)} = -\frac{e^{\frac{t}{a_\lambda}+X}}{\sqrt{X'^2 a_\lambda{}^2 e^{2(\frac{t}{a_\lambda}+X)} - t'^2}}(-\dot{X}'a_\lambda t'^2 + X'a_\lambda\dot{t}'t' + X'^2 a_\lambda t'' - \tag{54}$$

$$-a_\lambda t'X'X'' + X'^3 a_\lambda{}^2 e^{2(\frac{t}{a_\lambda}+X)} - X'^2 a_\lambda t' - 2t'^2 X'),$$

Equality of right-hand sides of (53) and (54) with (52) gives us:

$$(X'^2-1)(1-t'X'\dot{X}-t'X'^3+X') + e^{\frac{t}{a_\lambda}+X}(X'^2-1)^{\frac{3}{2}}(X'-X'^2-i\dot{X}') + \tag{55}$$

$$+(-\dot{X}'t'-X'X'') = 0$$

The two possibilities exist in this case:
the first is

$$X'^2 = 1, \tag{56}$$

or the second one reads

$$\begin{cases} 1 - t'X'\dot{X} - t'X'^3 + X' = 0, \\ X' - X'^2 - i\dot{X}' = 0, \\ \dot{X}'t' + X'X'' = 0. \end{cases} \tag{57}$$

But (57) is incompatible system. So, we conclude, that $X = X(R) = R + const.$ and from (52) $t = t(\tau)$.

### Matching Stephani and de Sitter solutions in the open case

The de Sitter and the Stephani metrics in open case we take in the form:

$$dS_{ds}{}^2 = dt^2 - \frac{a_\lambda{}^2 \sinh^2\left(\frac{t}{a_\lambda}\right)}{\sinh^2 X}(dX^2 + d\sigma^2), \tag{58}$$

$$dS_{st}{}^2 = \frac{\left(\frac{1}{2}\frac{\dot{\zeta}}{\zeta} + \coth(R+\tau)\right)^2}{\psi^2}d\tau^2 + \frac{1}{\zeta\sinh^2(R+\tau)}(dR^2 + d\sigma^2), \tag{59}$$

$t = t(\tau, R),\ X = X(\tau, R).$

For the Stephani solution normal vector has one non-zero component on the hypersurface $\tau = const.$:

$$\nu_0 = \frac{\frac{1}{2}\frac{\dot{\zeta}}{\zeta} + \coth(R+\tau)}{\psi}. \tag{60}$$

The non-zero components of the normal vector for the de Sitter solution are:

$$\nu_0 = \frac{X'a_\lambda \sinh\frac{t}{a_\lambda}}{\sqrt{X'^2 a_\lambda{}^2 \sinh^2\frac{t}{a_\lambda} - t'^2 \sinh^2 X}}, \tag{61}$$

$$\nu_1 = -\frac{t'a_\lambda \sinh\frac{t}{a_\lambda}}{\sqrt{X'^2 a_\lambda{}^2 \sinh^2\frac{t}{a_\lambda} - t'^2 \sinh^2 X}}. \tag{62}$$

From the equality of the first fundamental forms, next conditions follow:

$$\frac{1}{\zeta \sinh^2(R+\tau)} = X'^2 a_\lambda{}^2 \frac{\sinh^2 \dfrac{t}{a_\lambda}}{\sinh^2 X} - t'^2, \tag{63}$$

$$\frac{1}{\zeta \sinh^2(R+\tau)} = a_\lambda{}^2 \frac{\sinh^2 \dfrac{t}{a_\lambda}}{\sinh^2 X}. \tag{64}$$

From the equality of the right-hand sides of (63) and (64):

$$t'^2 = a_\lambda{}^2 \frac{\sinh^2 \dfrac{t}{a_\lambda}}{\sinh^2 X}(X'^2 - 1). \tag{65}$$

From the equality of the second fundamental forms, next conditions follow:

$$\frac{\psi}{\zeta \sinh^2(R+\tau)} = \frac{a_\lambda \sinh \dfrac{t}{a_\lambda}}{\sinh^2 X \sqrt{X'^2 a_\lambda{}^2 \sinh^2 \dfrac{t}{a_\lambda} - t'^2 \sinh^2 X}} \times$$

$$\times (t' \cosh X \sinh X - aX' \sinh \frac{t}{a_\lambda} \cosh \frac{t}{a_\lambda}),$$

$$\frac{\psi}{\zeta \sinh^2(R+\tau)} = \frac{1}{\sqrt{X'^2 a_\lambda{}^2 \sinh^2 \dfrac{t}{a_\lambda} - t'^2 \sinh^2 X}}(a_\lambda \dot{X} t'^2 \sinh \frac{t}{a_\lambda} -$$

$$\begin{aligned}
(66) \qquad &- a_\lambda \dot{t}' X t' \sinh \frac{t}{a_\lambda} + X \dot{t}'^2 \cosh \frac{t}{a_\lambda} - \frac{a_\lambda t' X'^2 \cosh X \sinh \dfrac{t}{a_\lambda}}{\sinh X} - \\[2mm]
&- a_\lambda X'^2 t'' \sinh \frac{t}{a_\lambda} + a_\lambda X'' \dot{t}' X' \sinh \frac{t}{a_\lambda} - \frac{a_\lambda{}^2 X'^3 \cosh \dfrac{t}{a_\lambda} \sinh^2 \dfrac{t}{a_\lambda}}{\sinh^2 X} + \\[2mm]
&+ X' t'^2 \cosh \frac{t}{a_\lambda}).
\end{aligned} \tag{67}$$

The equality of the right-hand sides of (66) and (67) gives:

$$\cosh \frac{t}{a_\lambda}(X'^2 - 1)^{\frac{3}{2}}(X' - X'^2 - \dot{i}X') + \sinh X(-X''X' - \dot{X}\dot{t}') +$$

$$+ \cosh X(X'^2 - 1)(-1 - X'^2) + \cosh X(X'^2 - 1)^{\frac{3}{2}}(X'^3 + \dot{X}X\dot{t}') = 0. \tag{68}$$

Two possibilities exist for satisfy this equation:

1)

$$X'^2 = 1, \tag{69}$$

or

2)

$$\begin{cases} X' - X'^2 - iX' = 0, \\ X''X' + \dot{X}\dot{t}' = 0, \\ 1 + X'^2 = 0, \\ X'^3 + \dot{X}X\dot{t}' = 0. \end{cases} \tag{70}$$

The last set of equation is an incompatible system, so, we conclude from (69) and (65), that $X = X(R) = R + const.$ and $t = t(\tau)$.

### Matching Stephani and de Sitter solutions in the closed case

The de Sitter and Stephani metrics we take in a more convenient for our purpose form:

$$dS_{ds}^2 = dt^2 - a_\lambda^2 \frac{\cosh^2 \frac{t}{a_\lambda^2}}{\cosh^2 X} (dX^2 + d\sigma^2), \tag{71}$$

$$dS_{st}^2 = \frac{\left(\frac{1}{2}\frac{\dot{\zeta}}{\zeta} + \tanh(R+\tau)\right)^2}{\psi^2} d\tau^2 - \frac{1}{\zeta \cosh^2(R+\tau)}(dR^2 + d\sigma^2), \tag{72}$$

We match on the hypersurface $\tau = const.$ $t = t(\tau, R), \ X = X(\tau, R)$. Dots and primes mean derivatives with respect to time and radial coordinate, respectively.

Non zero component for normal vector in the Stephani case on the matching hypersurface is:

$$\nu_0 = \frac{\frac{1}{2}\frac{\dot{\zeta}}{\zeta} + \tanh(R+\tau)}{\psi}. \tag{73}$$

And non zero components for the de Sitter case are:

$$\nu_0 = \frac{X' a_\lambda \cosh\frac{t}{a_\lambda}}{\sqrt{X'^2 a_\lambda^2 \cosh^2 \frac{t}{a_\lambda} - t'^2 \cosh^2 X}}, \tag{74}$$

$$\nu_1 = -\frac{t' a_\lambda \cosh\frac{t}{a_\lambda}}{\sqrt{X'^2 a_\lambda^2 \cosh^2 \frac{t}{a_\lambda} - t'^2 \cosh^2 X}}. \tag{75}$$

From the equality of the first fundamental forms, next conditions follow:

$$\frac{1}{\zeta \cosh^2(R+\tau)} = t'^2 - X'^2 a_\lambda^2 \frac{\cosh^2 \frac{t}{a_\lambda}}{\cosh^2 X}, \tag{76}$$

$$\frac{1}{\zeta \cosh^2(R+\tau)} = -\frac{a_\lambda^2 \cosh^2 \frac{t}{a_\lambda}}{\cosh^2 X}, \tag{77}$$

From the equality of the right-hand sides of (76) and (77):

$$t'^2 = \frac{a_\lambda{}^2 \cosh^2 \dfrac{t}{a_\lambda}}{\cosh^2 X}(X'^2 - 1). \tag{78}$$

From the equality of the second fundamental forms, next conditions follow:

$$\frac{\psi}{\zeta \cosh^2 (R + \tau)} = \frac{a_\lambda \cosh \dfrac{t}{a_\lambda}}{\cosh^2 X \sqrt{X'^2 a_\lambda{}^2 \cosh^2 \dfrac{t}{a_\lambda} - t'^2 \cosh^2 X}} \times$$

$$\times (a_\lambda X' \cosh \frac{t}{a_\lambda} \sinh \frac{t}{a_\lambda} - t' \sinh X \cosh X), \tag{79}$$

$$\frac{\psi}{\zeta \cosh^2 (R + \tau)} = \frac{1}{\sqrt{X'^2 a_\lambda{}^2 \cosh^2 \dfrac{t}{a_\lambda} - t'^2 \cosh^2 X}}(a_\lambda t' \cosh \frac{t}{a_\lambda} \times$$

$$\times (\dot{t}'X' - t'\dot{X}') + a_\lambda X' \cosh \frac{t}{a_\lambda}(t''X' - X''t') + X' \sinh \frac{t}{a_\lambda}\frac{1}{\cosh^2 X} \times$$

$$\times (X'^2 a_\lambda{}^2 \cosh^2 \frac{t}{a_\lambda} - t'^2 \cosh^2 X) + t'X'^2 a_\lambda \cosh \frac{t}{a_\lambda} \tanh X -$$

$$- X'\dot{t}'^2 \sinh \frac{t}{a_\lambda}). \tag{80}$$

Equality of the right parts of (79) and (80) gives:

$$\cosh X (t'\dot{X}' + X''X') + \sinh X (X'^2 - 1)(1 + X'^2 - X'^3 - X\dot{X}t') +$$

$$+ X' \sinh \frac{t}{a_\lambda}(X'^2 - 1)^{\frac{3}{2}}(\dot{t} + X' - 1) = 0. \tag{81}$$

A two possibilities exist to satisfy this equation:
1)

$$X'^2 = 1, \tag{82}$$

or
2)

$$\begin{cases} t'\dot{X}' + X''X' = 0, \\ 1 + X'^2 - X'^3 - X\dot{X}t' = 0, \\ \dot{t} + X' - 1 = 0. \end{cases} \tag{83}$$

This system of equation is incompatible, so, we conclude from (82) and (78), that $X = X(R) = R + const.$ and $t = t(\tau)$.

## Conclusions

Matching conditions for the Stephani and the de Sitter solutions on hypersourface $\tau = const.$ in spherically symmetric case have been obtained ($\tau$ is a time coordinate of the Stephani solution). The coordinates of the de Sitter solution were taken in a general form as arbitrary functions depending on the Stephani's time and radial coordinate. Matching was done both for special cases (flat, open, closed) and for the general case, that not concretize type of curvature. From the matching conditions the equality of densities on the matching hypersurfase has been obtained. From (15) a bound of the pressure exists. Also it was obtained, that de Sitter radial coordinate distinguishes from Stephani one on some shift and de Sitter time is a some arbitrary function depending on Stephani's time.